\documentclass[12pt]{iopart}

\usepackage[T1]{fontenc}
\usepackage{color}

\begin{document}

\title{Constant conditional entropy and related hypotheses}

\author{Ramon Ferrer-i-Cancho$^{1}$,  {\L}ukasz D\k{e}bowski$^{2}$ and Ferm\'{i}n Moscoso del Prado Mart\'{i}n$^3$} 

\address{$^1$ Complexity \& Quantitative Linguistics Lab \\
Departament de Llenguatges i Sistemes Inform\`atics, \\
TALP Research Center, Universitat Polit\`ecnica de Catalunya, \\
Campus Nord, Edifici Omega Jordi Girona Salgado 1-3. \\
08034 Barcelona, Catalonia (Spain)}

\address{$^2$ Institute of Computer Science \\
Polish Academy of Sciences \\
ul. Jana Kazimierza 5 \\
01-248 Warszawa, Poland}

\address{$^3$ Department of Linguistics \\ University of California at Santa Barbara \\ Santa Barbara, CA 93106-3100, USA} 

\eads{\mailto{rferrericancho@lsi.upc.edu}, \mailto{ldebowsk@ipipan.waw.pl} and \mailto{fermin.moscoso-del-prado@gmail.com}}

\begin{abstract}
Constant entropy rate (conditional entropies must remain constant as the sequence length increases) and uniform information density (conditional probabilities must remain constant as the sequence length increases) are two information theoretic principles that are argued to underlie a wide range of linguistic phenomena. Here we revise the predictions of these  principles to the light of Hilberg's law on the scaling of conditional entropy in language and related laws. We show that constant entropy rate (CER) and two interpretations for uniform information density (UID), full UID and strong UID, are inconsistent with these laws. Strong UID implies CER but the reverse is not true. Full UID, a particular case of UID, leads to costly uncorrelated sequences that are totally unrealistic. We conclude that CER and its particular cases are incomplete hypotheses about the scaling of conditional entropies.
\end{abstract}

\noindent{\it Keywords\/}: constant entropy rate, uniform information density, Hilberg's law.

\pacs{89.70.-a Information and communication theory  \\ 89.75.Da Systems obeying scaling laws \\ 05.40.-a Fluctuation phenomena, random processes, noise, and Brownian motion \\ 02.30.Lt Sequences, series, and summability}


\maketitle

\section{Introduction} 

\label{introduction_section}

Uniform information density and constant entropy rate
\cite{Levy2007a,Genzel2002a} are two information theoretic principles
that have been put forward to explain various linguistic phenomena,
e.g., syntactic reduction \cite{Levy2007a,Jaeger2010a} and the
frequency of word orders \cite{Maurits2010a}.  In order to present
these principles, we provide some definitions.  Formally, $u$ is a
linguistic sequence of $n$ elements, e.g., words or letters, i.e. $u =
x_1,...,x_n$, where $x_i$ is the $i$-th elements of the
sequence. $U_i$ is defined as the set of elements that can appear in
the $i$-th position of the sequence and $U$ as the set of all possible
sequences of length $n$, i.e. $U \subseteq U_1 \times ... ... \times
U_n$. Both $U$ and all the $U_i$'s are support sets, i.e. all their
members have non-zero probability and all non-members have zero
probability. $X_i$ is the random variable that takes values from the
set of elements $U_i$. Hereafter, $x_i$ stands for an element of $U_i$
or a value of $X_i$ by default.

The constant entropy rate (CER) hypothesis states that the conditional
entropy of an element given the previous elements remains constant
\cite{Genzel2002a}. To express it formally, $H(X_i|X_1,...,X_{i-1})$
is defined as the Shannon conditional entropy of the $i$-th element of
a sequence given the $i-1$ preceding elements. The CER states that
$H(X_i|X_1,...,X_{i-1})$ remains constant as $i$ increases ($i = 1, 2,
..., n$), i.e.
\begin{equation}
 H(X_1) = H(X_2|X_1) = ... = H(X_n | X_1,...,X_{n-1}). \label{constant_entropy_rate_hypothesis_equation}
\end{equation}
The uniform information density (UID) hypothesis \cite{Levy2007a, Jaeger2010a} states that the conditional probability of an element given the previous elements should remain constant.
To express it formally, $p(x_i | x_1,...,x_{i-1})$ is defined as the probability of $x_i$ given the preceding elements in $u$. 
We say than a particular utterance $u = x_1,...,x_n$ satisfies the UID condition if and only if 
\begin{equation}
p(x_1) = p(x_2|x_1) = ... = p(x_n | x_1,...,x_{n-1}).
\label{uniform_information_density_hypothesis_equation}
\end{equation}

Here we will review the validity of CER and UID to the
  light of the real scaling of the conditional and other entropies as
  $n$ increases, or equivalently, as $i$, the length of the prefix,
  increases
  \cite{Shannon1951a,Hilberg1990a,Ebeling1994,Schmitt1997a,Rao2010a}. We
  will show that the scaling of these entropies is inconsistent with
  CER and two interpretations of UID, strong UID and full UID, a
  particular case of strong UID. In essence, our arguments are the following.
  First, CER is inconsistent with Hilberg's law, a hypothesis on the
  entropy of natural language made on the base of Shannon's famous 
  experiment \cite{Shannon1951a}. Hilberg's law states that $H(X_n |
  X_1,...,X_{n-1}) \sim n^{-\beta}$, with $\beta \approx 1/2$
  \cite{Hilberg1990a}, whereas CER means $\beta = 0$.  Second, strong
  UID, i.e., all the utterances in $U$ satisfy the UID condition, is a
  particular case of CER. The latter can be easily shown. By taking
  logarithms and multiplying by $p(u)=p(x_1,...,x_n)$ on the UID
  condition (Eq.\
  \ref{uniform_information_density_hypothesis_equation}), the strong
  UID can be written equivalently as
\begin{eqnarray*}
p(u) \log p(x_1) & = & p(u) \log p(x_2|x_1) =  ... \\
       & = & p(u) \log p(x_n | x_1,...,x_{n-1})
\end{eqnarray*} 
for any $u = x_1, ..., x_n \in U$.
Summing over all $u \in U$ (the utterances that do not belong to $U$ have no effect thanks to the convention $0 \log 0 = 0$ \cite{Cover2006a}) and inverting the sign, the strong UID leads to
\begin{eqnarray*}
- \sum_u p(u) \log p(x_1) & = & - \sum_u p(u) \log p(x_2|x_1)  =  ... \\
       & = & - \sum_u p(u) \log p(x_n | x_1,...,x_{n-1}),
\end{eqnarray*}
where each part corresponds to the definition of a Shannon
(conditional) entropy, i.e. the definition of CER in Eq.\
\ref{constant_entropy_rate_hypothesis_equation} is recovered.

These ideas are developed in the coming sections,
  which will not only examine the meaning of strong UID but also that
  of full UID, a particularly degenerated version of real language
  where sequences of symbols are uncorrelated and entropies are
  maximum.

\section{The uniform information density hypothesis}

\label{mathematical_preliminaries_section}

First, let us inspect some consequences of the UID hypothesis.
One of the major challenges of the uniform density hypothesis is defining a criterion for the applicability of the hypothesis. UID was originally defined on a single sequence \cite{Levy2007a}. The utility and power of the hypothesis depends on its scope: the more sequences UID concerns, the better. We start with a very ambitious UID hypothesis, namely that UID holds for any sequence in $U$ that can be formed combining the elements of $U_i$, i.e. 
\begin{equation}
U =  U_1 \times ... \times U_n.
\label{full_UID_condition_equation}
\end{equation}
We call it full UID. We also consider a weaker but still strong
version, where UID holds also for any sequence in $U$ but
Eq.\ \ref{full_UID_condition_equation} does not need to be
satisfied. This version is called strong UID. In fact,
  full UID implies strong UID but the reverse is not true. To see the
  latter, consider support set $U=\{(a,b), (a,c), (d, e), (d,f)\}$,
  where $p(X_2 = x_2 | X_1 = x_1) = p(X_1 = x_1) = 1/2$ for any $(x_1,
  x_2)\in U$. We thus have strong UID but the full UID fails to hold
  since $U \neq U_1 \times U_2 = \{a,d\} \times
  \{b,c,e,f\}$.

The UID condition can be written in terms of the joint
probability. By the chain rule of conditional probability
\begin{equation}
p_u = p(x_1) p(x_2|x_1)...p(x_n | x_1,...,x_{n-1}).
\label{chain_rule_of_joint_probability_equation}
\end{equation}
and Eq.\ \ref{uniform_information_density_hypothesis_equation} we obtain that the UID hypothesis implies
\begin{equation}
p(x_1,...,x_n) = p(x_1)^n. \label{nex_UID_equation}
\end{equation}
That is, strong UID means that all sequences beginning with the same word are equally likely. Furthermore, noting that, by definition, we have 
\begin{equation}
p(x_i) = \sum_{x_1,...,x_{i-1}, x_{i+1},...,x_n} p(x_1, ..., x_n) \label{marginal_probability_equation}
\end{equation}
and applying Eq.\ \ref{nex_UID_equation}, we obtain
\begin{equation*}
p(x_1) = p(x_1)^n \sum_{\begin{array}{c} x_2, ..., x_n, \\ (x_1,...,x_n) \in U \end{array}} 1 = p(x_1)^n |U(X_1 = x_1)|
\end{equation*} 
where $U(X_1 = x_1)$ is the subset of $U$ containing all the sequences
where $X_1 = x_1$.  Therefore,
\begin{equation}
p(x_1) = |U(X_1 = x_1)|^\frac{1}{1-n}.
\label{reciprocal_of_natural_equation}
\end{equation}
Taking in particular $n=2$, we obtain that $p(x_1)$
  must be a reciprocal of a natural number.

\subsection{Full UID}

With the help of the properties of the UID hypothesis above it is easy
to show that full UID, implies, that for any sequence $x_1, ..., x_n$,
\begin{enumerate}
\item $x_1, ..., x_n$ are independent, i.e., 
\begin{equation*}
p(x_1, ..., x_n) = \prod_{i=1}^n p(x_i).
\end{equation*}
\item
The sets of elements that can appear at each position of the sequence have the same cardinality, i.e. $|U_1| = ... = |U_n|$.
\item
All words occurring in the same position are equally likely, i.e. $p(x_i) = 1/|U_1|$. 
\end{enumerate}
If UID holds for any sequence beginning with $x_1$ that can be formed
by combining elements from $U_2,...,U_n$, then $|U(X_1 = x_1)| =
\prod_{i=2}^n |U_i|$ and Eq.\ \ref{reciprocal_of_natural_equation}
becomes
\begin{equation}
p(x_1) = \left[\prod_{i=2}^n |U_i|\right]^\frac{1}{1-n}.
\label{p1_equation}
\end{equation} 
for $n\geq 2$. Eq.\ \ref{p1_equation} indicates that $p(x_1)$ is the same for any $x_1 \in U_1$. Thus, the condition 
\begin{equation*}
\sum_{x_1\in X_1} p(x_1) = 1
\end{equation*}
gives $p(x_1)=1/|U_1|$
and the UID condition in Eq.\ \ref{nex_UID_equation} becomes
\begin{equation}
p(x_1, ..., x_n) = |U_1|^{-n}.
\label{new4_UID_equation}
\end{equation}
Now we will derive $p(x_i)$ for $i\geq 2$. Employing Eq.\ \ref{new4_UID_equation}, Eq.\ \ref{marginal_probability_equation} 
can be written as (assuming $i\geq 2$)
\begin{equation*}
p(x_i) = |U_1|^{1-n} \frac{\prod_{j=2}^n |U_j|}{|U_i|} = |U_1|^{1-n} \prod_{j=2}^{i-1} |U_j| \prod_{j=i+1}^{n} |U_j|.
\end{equation*}
Notice that, again, $p(x_i)$ is the same for any $x_i \in X_i$.
Thus, the condition 
\begin{equation*}
\sum_{x_i\in X_i} p(x_i) = 1
\end{equation*}
gives $p(x_i)=1/|U_i|$ for any $i = 1,...,n$.
Now, we will show that $|U_i| = |U_1|$ for any $i = 1,...,n$. By definition, we have
\begin{equation*}
p(x_1,...,x_{i-1}) = \sum_{x_i} p(x_1,...,x_i),
\end{equation*}
which, thanks to UID (recall Eq.\ \ref{new4_UID_equation}), becomes
\begin{equation*}
p(x_1,...,x_{i-1}) = |U_1|^{-i}\sum_{x_i\in U_i} 1 = |U_1|^{-i} |U_i|. \\
\end{equation*}
Applying UID again (Eq.\ \ref{new4_UID_equation}) to the l.h.s, gives $|U_1|^{1-i} = |U_1|^{-i} |U_i|$,
an thus $|U_1| = |U_i|$ and $p(x_i) = 1/|U_1|$ for any $i = 1,...,n$.
Now it is easy to show that $x_1,...,x_n$ are independent. The definition of independence, i.e.
$p(x_1, ..., x_n) = \prod_{i=1}^n p(x_i),$
holds trivially under UID since then $p(x_1, ..., x_n) =
|U_i|^n$ and $p(x_i) = 1/|U_i|$ for any $i = 1,...,n$.
Finally, notice that full UID means that $p(X_i = x_i) = 1/|U_1|$ but does not imply that the elements of the sequences (regardless of their position) are  equally likely. If all the sequences of $U=\{(a,a), (a,c), (b,a), (b,c)\}$ have probability $1/4$, one then has full UID but the probability of producing $a$ is $4/8$ while the probability of producing $c$ is $2/8$.

\subsection{The relationship between CER and strong UID}

Unravelling the relationship between CER and UID is
also in need. For instance, \cite{Maurits2010a} is
based on the idea of UID but its mathematical implementation is based
on the definition of CER. Strong UID implies CER (Section \ref{introduction_section}) but it will be shown that the reverse implication does not hold. Consider CER with
  $n = 2$, i.e. $H(X_2|X_1) = H(X_1)$, and independence between $X_1$
  and $X_2$, i.e. $H(X_2 | X_1) = H(X_2)$. Thus, $H(X_1) =
  H(X_2)$. Assume also that $U_1 = \{a,b\}$ and $U_2 = \{c,d\}$. If one
  has $p(X_1 = a) = p(X_2 = c) = 2/5$, $p(X_1 = b) = p(X_2 = d) =
  3/5$, then one has CER but strong UID does not hold because $2/5$ is
  not the reciprocal of a natural number, a condition for strong UID
  noticed beforehand.

\section{The real scaling of entropies versus constant entropy rate}

\label{rejecting_CER_section}

A serious consequence of the properties of full UID is that it is
totally unrealistic with respect to natural language for several
reasons.  First, full UID leads to a sequence of independent elements,
while long range correlations pervade linguistic sequences both at the
level of letters and the level of words,
e.g. \cite{Montemurro2001b,Ebeling1994,Alvarez2006a,Moscoso2011a,Altmann2012a}. Second, full UID
is problematic because entropies are maximum. $H(X_i)$ is maximum for
any $i = 1,...,n$ as $H(X_i) = \log |U_i|$.  The joint entropy is
also maximum because \cite{Cover2006a}
\begin{equation}
H(X_1, ..., X_n) \leq \sum_{i=1}^n H(X_i)
\label{upper_bound_of_joint_entropy_equation}
\end{equation}
in general but full UID transforms the inequality of Eq. \ref{upper_bound_of_joint_entropy_equation} into a mere equality 
because the elements making a sequence are independent. Since entropy
is a measure of cognitive cost \cite{Moscoso2004a,Ferrer2007a}, full
UID means the entropy related costs are maximum.

Last but not least, the plausibility of UID or CER for natural
language is undermined by the results of celebrated experiments. Let
$H_e(X_n | X_1,...,X_{n-1})$ be an estimate of $H(X_n |
X_1,...,X_{n-1})$ from real data and $\varepsilon_n$ the error of the
estimate, i.e.,
\begin{equation*}
H_e(X_n | X_1,...,X_{n-1}) = H(X_n | X_1,...,X_{n-1}) + \varepsilon_n,
\end{equation*}
where $\varepsilon_n\ge 0$ in general (by the nonegativity of
Kullback-Leibler divergence and Kraft inequality, it follows that the
errors $\varepsilon_n$ are positive if entropy is estimated by means
of universal probability or universal coding
\cite{Cover2006a}). Hilberg \cite{Hilberg1990a} reanalyzed Shannon's
estimates of conditional entropy for English \cite{Shannon1951a} and
discovered that
\begin{equation}
H_e(X_n | X_1,...,X_{n-1}) \approx C_e n^{\alpha-1}
\label{hilbergs_law}
\end{equation}
with $C_e>0$, $\alpha \approx 0.5$, and $n\leq 100$
characters.
Extrapolating Hilberg's law, Eq.\ \ref{hilbergs_law}, for $n\gg 100$
requires some caution. If accepted with $\varepsilon_n=0$, Eq.\
\ref{hilbergs_law} would imply asymptotic determinism of human
utterances with an entropy rate $h=\lim_{n\rightarrow\infty} H(X_n |
X_1,...,X_{n-1})$ equal to $0$. 
Thus it is more plausible to accept
that the scaling law of the true entropy (not its
  estimate) is
\begin{equation}
H(X_n | X_1,...,X_{n-1}) \approx C n^{\alpha-1}+h,
\label{hilbergs_law_modified}
\end{equation}
with $C<C_e$ and a sufficiently small constant
$h>0$. That the modified Hilberg law, Eq.\
\ref{hilbergs_law_modified}, is indeed valid for natural language for
$n\gg 100$ characters can be corroborated by the following fact: the
modified Hilberg's law implies a lower bound for the growth of $V$,
the observed vocabulary size, as a function of the $T$, the text
length. Namely, Eq.\ \ref{hilbergs_law_modified} implies that $V$
grows at least as $\sim T^{\alpha}/\log T$ \cite{Debowski2011a}, which
is in good accordance with the real growth of $V$ \cite{Herdan1964a}.

In contrast, CER is a competing hypothesis on $H(X_n |
  X_1,...,X_{n-1})$, the true conditional entropy.  Assuming that
CER, Eq.\ \ref{constant_entropy_rate_hypothesis_equation}, holds in
spite of Hilberg's law, Eq.\ \ref{hilbergs_law}, is equivalent to
stating that the errors $\varepsilon_n \approx C_e n^{\alpha-1}-H(X_1)$
are systematically decreasing as $n$ increases and negative for
moderate $n$ if $H(X_1)$ is large enough. This seems
unrealistic since, as we have stated above, the
errors of entropy estimates should be positive in general
and it is unlikely that the errors are systematically
  diminishing (undersampling, a very important source of error, usually increases as $n$ increases).

The disagreement between real language and CER concerning the decay of
conditional entropy can be rephrased in terms of
other entropic measures: $H(X_1,...,X_n)$ the joint or block entropy
\cite{Schmitt1997a,Rao2010a} and $H(X_1, ...,X_n)/n$, the joint entropy per unit
\cite{Ebeling1994}.  It is easy to infer the scaling of these
entropies from the modified Hilberg law, Eq.\
  \ref{hilbergs_law_modified}, by means of the chain rule of the
joint entropy \cite{Cover2006a}, which yields
\begin{equation}
\fl H(X_1, ..., X_n) = \sum_{i=1}^n H(X_i | X_1,...,X_{i-1}) \approx \int_{0}^n [C m^{\alpha-1} + h] dm = \alpha^{-1} C n^{\alpha} + h n, \label{joint_entropy_law_equation}
\end{equation}
and thus 
\begin{equation}	
H(X_1, ..., X_n)/n \approx \alpha^{-1} C  n^ {\alpha - 1} + h. 
\label{joint_entropy_per_unit_law_equation}
\end{equation}
In contrast, CER predicts a linear growth of the joint entropy, i.e.
\begin{equation}
H(X_1,..., X_n)/n = H(X_1),
\label{linear_growth_of_joint_entropy_equation}
\end{equation}
which can be proven by applying the definition of CER in Eq.\
\ref{constant_entropy_rate_hypothesis_equation} to the chain rule of
joint entropy
\cite{Cover2006a}. 

As an independent confirmation of Shannon's research, according to
Ref.\ \cite{Ebeling1994}, the estimates of $H(X_1, ...,X_n)/n$ for
sequences of letters from an English novel show good agreement with
Eq.\ \ref{joint_entropy_per_unit_law_equation} with $\alpha =
0.5$. Furthermore, the estimates of
$H(X_1,...,X_n)$ grow sublinearly with $n$ not only for texts in
English but also for sequences from many other languages with
different kinds of units \cite{Rao2010a,Schmitt1997a}. CER predicts that $H(X_1, ...,X_n)/n$ is
constant, Eq.\ \ref{linear_growth_of_joint_entropy_equation}, a
strikingly different result.
Therefore, the inconsistencies of CER are robust in the sense that do
not depend on entropic measure, the language or the units of the
sequence being considered.  The same inconsistencies concern full and
strong UID as it has been shown in Sections \ref{introduction_section} and 
\ref{mathematical_preliminaries_section} that $\mbox{Full UID}
\Rightarrow \mbox{Strong UID} \Rightarrow \mbox{CER}$.

\section{Discussion}

We have shown that CER and the two interpretations of the UID
hypothesis (full UID and strong UID) are inconsistent with the scaling
law for entropy of natural language called Hilberg's law. 
Future research should address the challenge of what modifications of
UID/CER can be consistent with real language.  In order to save
UID/CER, we envisage that probabilities and conditional entropies for
real language stem from a conflict between principles: one acting
towards UID/CER and another acting against UID/CER.
A similar conflict between principles has been hypothesized for Zipf's
law for word frequencies, namely that the frequency of the $i$-th most
frequent word of a text is $\sim i^{-\tau}$ \cite{Zipf1949a}: the law
emerges from a conflict between two principles, minimization of the
entropy of words and maximization of the mutual information between
words and meanings where none of the principles is totally realized
\cite{Ferrer2004e,Prokopenko2010a}. Indeed, Refs.\
\cite{Debowski2011a,Debowski2011b} show that there is a close
relationship between Zipf's law and Hilberg's law so the conflict of
principles that leads to Zipf's law may be the same that prevents
UID/CER from the full realization. 

It is worth noting that there are simple information theoretic
principles that lead to UID/CER. For instance, minimization of
conditional entropy, leads to Eq.\
\ref{constant_entropy_rate_hypothesis_equation} with $H(X_1) = 0$.
Interestingly, entropy minimization can be easily justified as it
implies the minimization of cognitive cost
\cite{Moscoso2004a,Ferrer2007a} and is used to explain Zipf's law for
word frequencies \cite{Ferrer2004e,Prokopenko2010a}.  Clearly, this
conditional entropy minimization could not be acting
alone as its total realization implies that all possible sequences
have probability zero except one and therefore Hilberg's law
  could not hold.
As the force towards UID/CER is not
working alone, the nature of the
second factor in conflict must be clarified. The view
  of UID/CER as a principle in conflict means that all the currently
  available explanations of linguistic phenomena based upon UID/CER,
  e.g., \cite{Genzel2002a, Levy2007a,Maurits2010a}, are {\em a priori} incomplete.

\ack We are grateful to R. Levy, F. Jaeger, S. Piantadosi and
E. Gibson for helpful discussions.  This work was supported by the
grant {\em Iniciaci\'o i reincorporaci\'o a la recerca} from the
Universitat Polit\`ecnica de Catalunya and the grants BASMATI
(TIN2011-27479-C04-03) and OpenMT-2 (TIN2009-14675-C03) from the
Spanish Ministry of Science and Innovation (RFC).

\section*{References}

\bibliographystyle{unsrt}


\begin{thebibliography}{10}

\bibitem{Levy2007a}
R.~Levy and T.~F. Jaeger.
\newblock Speakers optimize information density through syntactic reduction.
\newblock {\em Proceedings of the Twentieth Annual Conference on Neural
  Information Processing Systems}, 2007.

\bibitem{Genzel2002a}
D.~Genzel and E.~Charniak.
\newblock Entropy rate constancy in text.
\newblock In {\em Proceedings of the 40th Annual Meeting of the Association for
  Computational Linguistics (ACL-02)}, pages 199--206, 2002.

\bibitem{Jaeger2010a}
T.~F. Jaeger.
\newblock Redundancy and reduction: Speakers manage syntactic information
  density.
\newblock {\em Cognitive Psychology}, 61(1):23 -- 62, 2010.

\bibitem{Maurits2010a}
L.~Maurits, A.~A.~Perfors, and D.~Navarro.
\newblock Why are some word orders more common than others? a uniform
  information density account.
\newblock In {\em Advances in Neural Information Processing Systems},
  volume~23, pages 1585--1593, 2010.

\bibitem{Shannon1951a}
C.~Shannon.
\newblock Prediction and entropy of printed {English}.
\newblock {\em Bell System Technical Journal}, 30:50--64, 1951.

\bibitem{Hilberg1990a}
W.~Hilberg.
\newblock {Der} bekannte {Grenzwert} der redundanzfreien {Information} in
  {Texten} — eine {Fehlinterpretation} der {Shannonschen} {Experimente}?
\newblock {\em Frequenz}, 44:243--248, 1990.

\bibitem{Ebeling1994}
W.~Ebeling and T.~P\"{o}schel.
\newblock Entropy and long-range correlations in literary {English}.
\newblock {\em Europhysics Letters}, 26(4):241--246, 1994.

\bibitem{Schmitt1997a}
A.~O. Schmitt and H.~Herzel.
\newblock Estimating the entropy of {DNA} sequences.
\newblock {\em Journal of Theoretical Biology}, 188(3):369 -- 377, 1997.

\bibitem{Rao2010a}
R.~Rao.
\newblock Probabilistic analysis of an ancient undeciphered script.
\newblock {\em IEEE Computer}, 43:76--80, 2010.

\bibitem{Cover2006a}
T.~M. Cover and J.~A. Thomas.
\newblock {\em Elements of information theory}.
\newblock Wiley, New York, 2006.
\newblock 2nd edition.

\bibitem{Montemurro2001b}
M.~Montemurro and P.~A. Pury.
\newblock Long-range fractal correlations in literary corpora.
\newblock {\em Fractals}, 10:451--461, 2002.

\bibitem{Alvarez2006a}
E.~Alvarez-Lacalle, B.~Dorow, J.-P. Eckmann, and E.~Moses.
\newblock Hierarchical structures induce long-range dynamical correlations in
  written texts.
\newblock {\em Proceedings of the National Academy of Sciences},
  103:7956--7961, 2006.

\bibitem{Moscoso2011a}
F.~Moscoso del Prado~Mart\'{\i}n.
\newblock The universal `shape' of human languages: spectral analysis beyond
  speech.
\newblock {\em PLoS ONE}, page in press, 2011.

\bibitem{Altmann2012a}
E.~A. Altmann, G.~Cristadoro, and M.~D. Esposti.
\newblock On the origin of long-range correlations in texts.
\newblock {\em Proc. Natl. Acad. Sci. USA}, 109:11582--11587, 2012.

\bibitem{Moscoso2004a}
F.~{Moscoso del Prado Mart{\'i}n}, Alexandar Kosti{\'c}, and {R. H. Baayen}.
\newblock Putting the bits together: an information theoretical perspective on
  morphological processing.
\newblock {\em Cognition}, 94:1--18, 2004.

\bibitem{Ferrer2007a}
R.~{Ferrer-i-Cancho} and A.~D\'iaz-Guilera.
\newblock The global minima of the communicative energy of natural
  communication systems.
\newblock {\em Journal of Statistical Mechanics}, page P06009, 2007.

\bibitem{Debowski2011a}
\L. D\k{e}bowski.
\newblock On the vocabulary of grammar-based codes and the logical consistency
  of texts.
\newblock {\em IEEE Transactions on Information Theory}, 57:4589--4599, 2011.

\bibitem{Herdan1964a}
G.~Herdan.
\newblock {\em Quantitative linguistics}.
\newblock Butterworths, London, 1964.

\bibitem{Zipf1949a}
G.~K. Zipf.
\newblock {\em Human behaviour and the principle of least effort}.
\newblock Addison-Wesley, Cambridge (MA), USA, 1949.

\bibitem{Ferrer2004e}
R.~{Ferrer i Cancho}.
\newblock {Zipf's} law from a communicative phase transition.
\newblock {\em European Physical Journal B}, 47:449--457, 2005.

\bibitem{Prokopenko2010a}
M.~Prokopenko, N.~Ay, O.~Obst, and D.~Polani.
\newblock Phase transitions in least-effort communications.
\newblock {\em J. Stat. Mech.}, page P11025, 2010.

\bibitem{Debowski2011b}
\L. D\k{e}bowski.
\newblock Excess entropy in natural language: present state and perspectives.
\newblock {\em Chaos}, 21(3):037105, 2011.

\end{thebibliography}

\end{document}